\documentclass[fleqn,usenatbib]{mnras}
\usepackage{graphicx}
\usepackage{amsmath,amssymb,amsfonts}
\usepackage[cal=cm]{mathalfa}
\DeclareRobustCommand{\VAN}[3]{#2}
\let\VANthebibliography\thebibliography
\def\thebibliography{\DeclareRobustCommand{\VAN}[3]{##3}\VANthebibliography}
\newcommand{\daa}{\Delta\alpha/\alpha}

\title[The Mystery of Alpha and the Isotopes]{The Mystery of Alpha and the Isotopes}

\author[J. K. Webb et al]{John K. Webb$^{1,2,3}$\thanks{jw978@cam.ac.uk},
Chung-Chi Lee$^3$,
Dinko Milakovi{\'c}$^{4,5}$,
Victor V. Flambaum$^6$,\newauthor
Vladimir A. Dzuba$^6$,
Jo\~ao Magueijo$^7$.\\\\
$^1${Institute of Astronomy, University of Cambridge, Madingley Road, Cambridge CB3 0HA, UK.}\\
$^2$Clare Hall, University of Cambridge, Herschel Rd, Cambridge CB3 9AL.\\
$^3$Big Questions Institute, Level 4, 55 Holt St., Surry Hills, Sydney, NSW 2010, Australia.\\
$^4$Institute for Fundamental Physics of the Universe, Via Beirut, 2, 34151 Trieste, Italy\\
$^5$INAF - Osservatorio Astronomico di Trieste, via Tiepolo 11, 34131, Trieste, Italy\\
$^6$School of Physics, University of New South Wales, Sydney, NSW 2052, Australia.\\
$^7$Theoretical Physics Group, The Blackett Laboratory, Imperial College, Prince Consort Rd., London, SW7 2BZ, UK.
}

\date{Accepted XXX. Received YYY; in original form ZZZ}
\pubyear{2024}
\begin{document}
\label{firstpage}
\pagerange{\pageref{firstpage}--\pageref{lastpage}}
\maketitle

\begin{abstract}
We report unbiased AI measurements of the fine structure constant $\alpha$ in two proximate absorption regions in the spectrum of the quasar HE0515$-$4414. The data are high resolution, high signal to noise, and laser frequency comb calibrated, obtained using the ESPRESSO spectrograph on the VLT. The high quality of the data and proximity of the regions motivate a differential comparison, exploring the possibility of spatial variations of fundamental constants, as predicted in some theories. We show that if the magnesium isotopic relative abundances are terrestrial, the fine structure constants in these two systems differ at the 7$\sigma$ level. A 3$\sigma$ discrepancy between the two measurements persists even for the extreme non-terrestrial case of 100\% $^{24}$Mg, if shared by both systems. However, if Mg isotopic abundances take independent values in these two proximate systems, one terrestrial, the other with no heavy isotopes, both can be reconciled with a terrestrial $\alpha$, and the discrepancy between the two measurements falls to 2$\sigma$. We cannot rule out other systematics that are unaccounted for in our study that could masquerade as a varying alpha signal. We discuss varying constant and varying isotope interpretations and resolutions to this conundrum for future high precision measurements.
\end{abstract}


%\quickwordcount{main}
%\detailtexcount{main}

\section{Introduction}

Increasingly high quality astronomical data gives persistent hints that the simple homogeneous and isotropic $\Lambda$CDM model may provide only an approximation to reality, and that a deeper understanding of the universe requires a more complex cosmological model e.g.\,\cite{Peebles2022, Abdalla2022}. This may involve a cosmology in which fundamental constants vary, either temporally or spatially \cite{Garćia-Berro2007, Uzan2011, Uzan2015, Martins2017}. The fine structure constant $\alpha = e^2/4\pi\epsilon_0\hbar c$ (SI) $= e^2/\hbar c$ (CGS) is testable in both senses, complementing other quantities such as the proton-to-electron mass ratio $\mu$, e.g. \cite{Flambaum2004}. A departure of $\alpha$ from its terrestrial value creates shifts in the observed frequencies $\omega_{obs}$ of atomic spectral lines compared to the laboratory value $\omega_{lab}$, given by $\omega_{obs} = \omega_{lab} + qx + O(x^2) \approx \omega_{lab} + 2q\daa$, 
where $x=(\alpha_{obs}/\alpha_{lab})-1$, $\daa = (\alpha_{obs} - \alpha_{lab})/\alpha_{lab}$, and $q$ is a sensitivity coefficient. Because $|q|$ increases strongly with nuclear charge, and because $q$ may take positive or negative values, comparing relative shifts between sets of atomic transitions having widely different $q$ values leads to strong constraints on any possible deviation in $\alpha_{obs}$ from the terrestrial value. This technique, the {\it Many Multiplet method} \citep{Dzuba1999, Webb1999}, provides a stringent probe for new physics and is a science driver for both the ESPRESSO instrument on the VLT \citep{Pepe2021} and the forthcoming ELT itself \citep{Marconi2016,Tamai2018}.

\section{Astronomical data}

The data used for this study is the ESPRESSO quasar spectrum of HE0515$-$4414, as analysed in \cite{Murphy2022,Lee2023bias, Webb2024}. HE0515$-$4414 is unusual in that its sightline intersects a particularly extensive system of narrow absorption lines at $z_{abs} = 1.15$ spanning $\sim$730 km/s \citep{Reimers1998}. Although a complex this large is rare, it is not unique; Q0013$-$004 (also catalogued as UM224, \cite{MacAlpine1977}) has an even more extensive system spanning at least $\sim$800 km/s \citep{Petitjean2002}. Detailed studies of the HE0515$-$4414 $z_{abs} = 1.15$ complex are reported in \cite{Quast2008, Milakovic2021, Murphy2022, Lee2023bias} and others. The data acquisition, laser frequency comb (LFC) wavelength calibration, and pipeline reduction procedures of the ESPRESSO spectrum are described in \cite{Murphy2022}. Various authors have sub-divided this absorption complex into isolated regions, each being flanked by sufficient continuum points to be treated independently. The advantages of doing this are that it simplifies the analysis and calculation times and, evidently, provides spatial information. In the analysis presented here, we partition the absorption complex into three regions (to reduce computation times and to allow independent $\daa$ measurements across the absorption complex) using the same boundaries as in \cite{Murphy2022, Lee2023bias, Webb2024}. The central region provides a very weak $\daa$ constraint so here we focus on the left and right regions only. The atomic transitions used to measure $\daa$ in both regions of the HE0515$-$4414 $z_{abs}=1.15$ system are tabulated and plotted in \cite{Murphy2022, Lee2023bias}. The left region comprises weaker, generally unsaturated absorption lines. We use 8 transitions from MgII, FeII, and MgI for the left region. For the right region, we use 19 transitions in 18 regions: (1) MgI 2026, (2) CrII 2056, (3) ZnII 2062 + CrII 2062, (4) CrII 2066, (5) FeII 2260, (6) FeII 2344, (7) FeII 2374, (8) FeII 2382, (9) MnII 2576, (10) FeII 2586, (11) MnII 2594, (12) FeII 2600, (13) MnII 2606, (14) MgII 2796, (15) MgII 2803, (16) MgI 2852, (17) FeI 2484, and (18) FeI 2523.

\section{Methodology}

Absorption profiles are modelled using compound broadening, defined by the turbulent contribution and the gas temperature $T$. It is important to use compound broadening \cite{Lee2020AI-VPFIT} rather than turbulent because (i) compound incorporates turbulent as a limiting case and is hence more general, (ii) it is well known that absorption system gas temperatures are typically $\sim$10$^4$K e.g. \cite{Carswell2012, Cooke2015, Welsh2020, Noterdaeme2021, Milakovic2021}, and (iii) empirically, compound requires fewer free parameters to fit an absorption complex \cite{Lee2023bias}, further suggesting it is the physically appropriate model to use. We constrain individual components to have Doppler parameters $0.3 \le b \le 10$ km/s. The lower limit was set to be below the smallest values found in practice and hence to avoid numerical difficulties caused by component rejection in the fitting procedure if the free parameter $b$ happens to reduce to zero (resulting in rejection of that component). The upper limit was chosen to avoid any possible artificially high $b$ values occurring in some models. The number of free model parameters are minimised by tieing physically related parameters \citep{web:VPFIT}. Specifically, for each redshift component, we tied component redshifts and component velocity dispersion parameters $b$. All column density parameters are free so that component to component ionization differences are accounted for i.e. the model building procedure allows for different velocity structures for different atomic species.

In addition to Voigt parameters, each data segment includes additional continuum parameters, the purpose of which is to fine-tune the user-provided continuum model \citep{web:VPFIT}. The covariance matrix then contains terms associated with those additional free parameters, such that error propagation for all parameters is handled correctly. Models are convolved with a Gaussian instrumental profile. We are aware that the ESPRESSO instrumental profile (IP) deviates slightly from Gaussianity but the true IP (a function of parameters such as position on detector and wavelength) has not yet been measured. We return to the potential impact of this approximation later.

Modelling is carried out using {\sc ai-vpfit}, which requires no human interactions, and hence is completely objective \citep{Lee2020AI-VPFIT}. Since model non-uniqueness may occur \citep{Lee2021}, we compute a total of 200 independent models covering both left and right regions (50 for each region, two Mg isotope settings). By ``independent'', we mean that during model construction, sequential placement of trial absorption components is random, using a different set of random number seeds for each model. The fine structure constant $\alpha$ is a free parameter throughout development of the absorption system kinematic structure. This is crucial to ensure bias-free $\daa$ measurements, \cite{Webb2022,Lee2023bias}. Model selection through the AI stages \cite{Lee2020AI-VPFIT} is provided by the {\it Spectroscopic Information Criterion}, SpIC (which, in this context, performs better than both the corrected Akaike information criterion and the Bayesian information criterion) \citep{Webb2021}. 

\begin{figure*}
\centering
\includegraphics[width=1.0\linewidth]{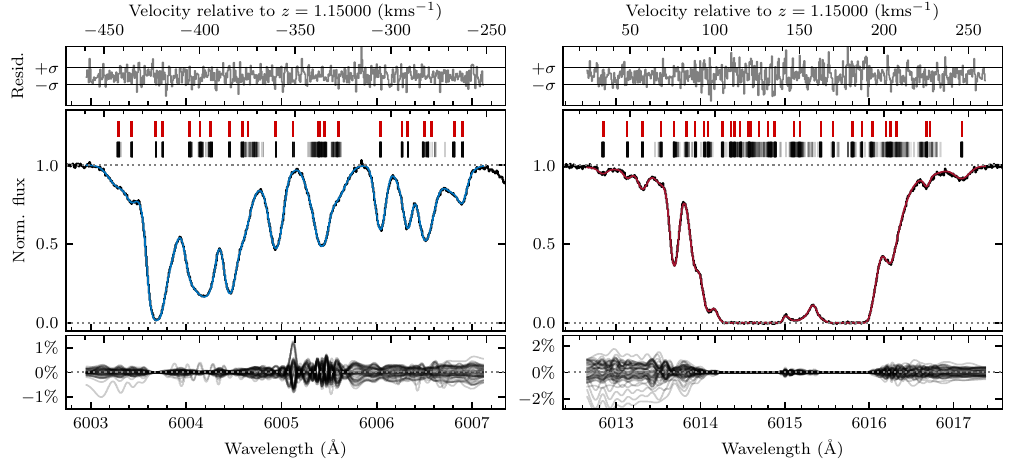}
\caption{Mean of 50 compound broadened models, terrestrial Mg isotopes. Middle panels: Mg{\sc ii} 2796 for left (blue continuous line) and right (red) regions. The black histogram under each mean model curve shows the data. The coloured tick marks immediately above the row of grey tick marks show one comparison model (the smallest SpIC value), illustrating the typical number of components required to obtain a good model. The grey tick marks show the absorption component positions for all 50 models. The two sets of tick marks provide a direct and powerful illustration of model non-uniqueness. Top panels: normalised residuals between mean model and data. Lower panels: percentage scatter between the mean model and each individual model, $(m_i - \left <m_i \right>)/\left< m_i \right>$. All 50 for each region are plotted.}
\label{fig:mean50}
\end{figure*}

\begin{figure}
\centering
\includegraphics[width=1.0\linewidth]{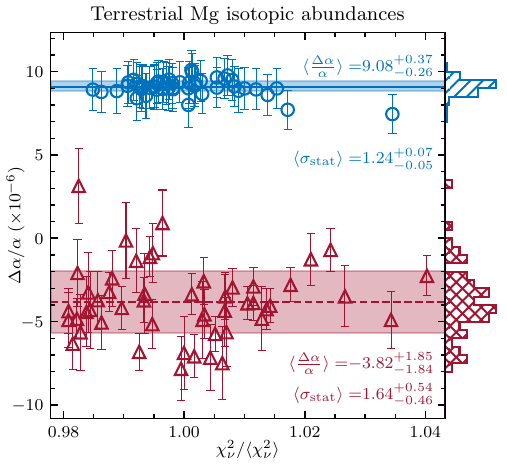}
\caption{{\sc ai-vpfit} measurements of $\alpha$ in the $z_{abs}=1.15$ complex using terrestrial relative isotopic abundances for Mg. Each point is an independently constructed model, 50 for each region. The abscissa is $\chi^2_{\nu}/\langle\chi^2_{\nu}\rangle$ rather than just $\chi^2_{\nu}$, for aesthetic display, since $\langle \chi^2_{\nu} \rangle_{left} \approx 0.78$ whilst $\langle \chi^2_{\nu} \rangle_{right} \approx 1.0$. Blue circles show the left region, red triangles show the right. The insets show the weighted mean $\daa$, the empirical non-uniqueness $\pm 1\sigma$ error (= $\pm 34$\% empirical scatter, also shown by the shaded regions), the ordinary mean of the statistical uncertainties (from the best-fit covariance matrix) and its $1\sigma$ error. The number of components of the primary species, averaged over all 50 models, is $21.98 \pm 0.77$ for the left region and $30.18 \pm  1.41$ for the right region.
%, so we claim that} the right region's uncertainty is dominated by non-uniqueness \mgt{rather than the left region.} \mgt{To John: (1) the number of components is one of the measurement of the non-uniqueness problem. An alternative, we may try to use $\sigma(N_{primary}) / N_{primary}$ as a measurement: $0.77/21.98 = 3.5\%$ (left) and $0.77/21.98 = 4.7\%$ (right) (2) I think the saturated profile at the right region is another source.}
}
\label{fig:terrestrialMg}
\end{figure}

\section{Results using terrestrial isotopic abundances}

The terrestrial Mg isotopic abundances used are 78.99, 10.00, 11.01\% for $^{24,25,26}$Mg respectively \citep{Asplund2009}. The {\sc ai-vpfit} $\daa$ results (2 spectral regions, 50 $\daa$ measurements for each) are illustrated in Figs.\,\ref{fig:mean50} and \ref{fig:terrestrialMg}. The process of calculating multiple AI models emulates having a set of independent human interactive model builders, each of whom would opt to do things in a slightly different way. All models satisfy the usual reduced $\chi^2$ statistical acceptance \citep{Lee2023bias}, each model is a valid representation of the data, and hence the set of models enable us to quantify the contribution of model non-uniqueness to the $\daa$ error budget \citep{Lee2021}.

Fig.\,\ref{fig:terrestrialMg} insets show the weighted mean $\daa$ with its intrinsic scatter, and the mean statistical (i.e.\,covariance matrix) error, with its scatter. $(\daa)_{left}$ exhibits $\sim$7$\times$ less scatter than $(\daa)_{right}$ i.e.\,model non-uniqueness is far more severe for the right region, and in fact {\it dominates} the overall $\daa$ error budget. The left region is more suited to a $\daa$  measurement than is the right. $(\daa)_{left}$, assuming terrestrial isotopic relative abundances, deviates from the terrestrial value at a significance level of $\sim$7$\sigma$. $(\daa)_{right}$ on the other hand, deviates from terrestrial by only $\sim$2.3$\sigma$.

\section{Results using non-terrestrial magnesium isotopes}

\begin{figure}
\centering
\includegraphics[width=1.0\linewidth]{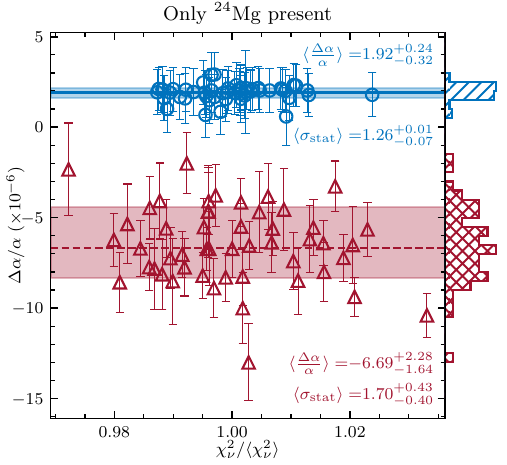}
\caption{Similar to Fig.\,\ref{fig:terrestrialMg}, except all {\sc ai-vpfit} models were fitted using 
100\% $^{24}$Mg. The number of components of the primary species, averaged over all 50 models, is $22.04 \pm 0.70$ for the left region and $30.32 \pm 1.42$ for the right region.}
\label{fig:AllMg24}
\end{figure}

If in fact the gas clouds we have analysed have non-terrestrial isotopic abundances, imposing terrestrial values in the models will create artificial line shifts, resulting in a biased $\alpha$ measurement \citep{Webb1999}. Given the lower atomic mass and hence wider isotopic and hyperfine component spacings, this problem is most severe for magnesium. The results of modelling the data using an extreme case of 100\% $^{24}$Mg{\sc ii} are shown in Fig.\,\ref{fig:AllMg24}. Only a few quasar absorption system measurements of the Mg isotopic ratios have been done and the constraints are not yet tight (and in any case might vary spatially), so the best we can do in the absence of good constraints (when using Mg), is to allow an extreme case (here, removing the heavier isotopes entirely) to examine the impact on $\daa$. Isotopic structure calculations and measurements are discussed in e.g.\,\cite{Ashenfelter2004PRL, Berengut2003, Kozlov2004, Berengut2005, Salumbides2006, Batteiger2011}. A comprehensive treatment of Galactic chemical evolution is given in \cite{Kobayashi2020}. Recent upper limits on the relative abundance of the heavy Mg isotopes is given in \cite{Noterdaeme2021}, and \cite{Murphy2014} provides a compilation of atomic data useful for Many Multiplet $\daa$ measurements. The atomic data used for all measurements here are as provided with {\sc vpfit} v12.4 \citep{web:VPFIT}. To explore the impact of non-terrestrial isotopic ratios using {\sc ai-vpfit}, it is a simple matter of modifying {\sc vpfit}'s {\it atom.dat} file accordingly, then running {\sc ai-vpfit} to obtain new model fits. One should {\it not} simply modify isotopic ratios and then re-minimise existing models that were established using terrestrial abundances, as that procedure has unpredictable bias.

Comparing the results illustrated in Figs.\,\ref{fig:terrestrialMg} and \ref{fig:AllMg24}, we see that the impact of using 100\% $^{24}$Mg rather than terrestrial is greater on the left region than the right; the former shifts by almost twice the latter. The likely reason for this is that the latter comprises many saturated Mg components, increasing some of the redshift uncertainties and hence permitting a greater range of solutions for $\daa$. Note that the Mg isotope column densities are not individually free parameters. The relative isotopic abundances are fixed in the input atomic data file. Therefore it is impossible that, in the results we present, there is any bias on alpha for the red (right) region associated with saturation {\it per se}.

One can see from either Fig.\,\ref{fig:terrestrialMg} or \ref{fig:AllMg24} that the spread in $\daa$ values is much larger for the right region than for the left. It is also worth noting that although the right region exhibits greater scatter, no evidence for multiple minima is seen. Fig.\,\ref{fig:mean50} shows that the right region comprises more absorption components than does the left and also that the right region is saturated (for Mg{\sc ii} 2796). Both of these effects indicate that line blending is more severe for the right region than for the left, and hence that both effects contribute to the increased scatter (or model non-uniqueness) for the right region. 

\section{Systematics}

(i) A recent study \citep{Schmidt2021} identified an unexplained high-frequency pattern with deviations up to 7 m/s in the ESPRESSO LFC calibration residuals. We can approximate the impact on $\daa$ as follows. We use Fe{\sc ii}2383 as an example, for which $q=1505$ cm$^{-1}$. Denote $y = \daa$, so $dy = d\omega_{obs}/2q$ and $d\omega_{obs} \approx 0.001$ cm$^{-1}$. For a 7 m/s local calibration fluctuation, the impact on $\daa$ is $dy \approx 3 \times 10^{-7}$. However, we fit 8 atomic transitions (left region) in this analysis which dilutes the effect to an upper limit of $\approx 1 \times 10^{-7}$, an order of magnitude below the $\daa$ statistical uncertainty. The effect is diminished yet further  because we have multiple velocity components in each absorption complex. The impact of this high-frequency calibration residuals pattern may thus be ignored in this case. (ii) a detector problem exists for ESPRESSO data: incomplete identification of warm pixels in raw frames leaves spurious spectral features in the extracted spectrum \citep{Pasquini2025}. Manual correction of the data during initial processing \citep{Murphy2022} largely removes these although a few unidentified weak features might remain and bias $\daa$. Nevertheless, it is worth noting that one of the huge advantages of our AI procedure is that the velocity structures obtained are completely unbiased. Further, the presence of weak unidentified interlopers is tested for (automatically) by AI-VPFIT. %None were found. 
(iii) We have focused only on the effect of Mg isotopes. Non-terrestrial Fe isotopes also impact, although to a lesser degree because the isotopic spacings are approximately half those of Mg, and the Fe{\sc ii} heavy element isotopes are approximately symmetric in wavelength about the strongest isotope, further diminishing the impact of Fe{\sc ii} isotope variations. (iv) Sensitivity coefficient $q$ uncertainties will bias the measured $\daa$ (unless the true $\daa=0$). (v) It is impossible to entirely preclude unidentified weak blends biasing any particular $\alpha$ measurement (even though {\sc ai-vpfit} will identify some (a search was carried out in \cite{Milakovic2021}). Conclusive spacetime variations (or lack thereof) can only be achieved with a statistical survey. (vi) Wavelength calibration made use of a Gaussian profile for centroiding LFC lines. Also, our {\sc ai-vpfit} models were convolved with the nominal (constant) Gaussian IP. Both approximations create an additional $\daa$ uncertainty that we cannot currently estimate. However, the two measurements made here are fairly close in wavelength and hence detector position, so we may reasonably hope the Gaussian approximation justifies the comparative $\daa$ measurements made in this analysis (an assumption that will be tested in future work when the true IP becomes available). (vii) Finally, our sub-division of the $z_{abs}=1.15$ complex into ``left'' and ``right'' was necessarily done on the basis of convenient separation by sufficient stretches of continuum \citep{Wilczynska2015}. This need not exactly match any true physical division.

\section{Comparison with previous measurements}

Previous measurements of $\daa$ in the $z_{abs}=1.15$ towards HE0515$-$4414 are tabulated in \cite{Murphy2022}. All previous analyses (other than \cite{Milakovic2021}) required interactive human decision making, producing a single model of the complex. {\it Any} analysis which holds $\daa=0$ during model development is likely to be heavily biased towards the terrestrial result, and is unlikely to reflect the true statistical measurement \citep{Webb2022, Lee2023bias, Webb2024}. The approach in this paper is entirely different. We have generated more than 200 independent AI models, allowing us to explore hitherto unknown systematics and biases. Such a task would previously have required 200 human beings, each working for many months, with inhomogeneous results.

The analysis in \cite{Milakovic2021} also made use of {\sc ai-vpfit}, but was based on only one AI model. Nevertheless, that detailed analysis attempted something we have not tried here and that had not been attempted before: measuring $\daa$ in individual velocity components across the whole absorption complex. The data used in \cite{Milakovic2021} have slightly lower spectral resolution ($R=115,000$ compared to $145,000$ here) and lower signal to noise ($\approx\!50$ per  0.75 km/s pixel at 6000{\AA}), so constraints in the earlier study are not as tight as those for the ESPRESSO data studied here (which has a signal to noise per 0.4 km/s pixel of $\approx\!105$ at 6000{\AA}). 

Evidently, uncertainties on individual components in \cite{Milakovic2021} are large, but interesting effects are nonetheless seen. The analysis in \cite{Milakovic2021} sub-divides the complex into 5 spectral regions. Regions I and II combined in \cite{Milakovic2021} correspond to our left region. Interestingly, although the telescope, instrument, data, and analysis are entirely independent, the high value of $\daa$ found here for the left region agrees well with \cite{Milakovic2021} (suggesting {\it instrumental} systematics do not dominate in both studies). Further, our right region corresponds to Region V in \cite{Milakovic2021} and again the agreement between \cite{Milakovic2021} and our measurement here is good. The inference from this is one of encouragement; overcoming systematics and obtaining experimental consistency are both achievable.

\section{Interpretation and discussion}

In future work, measurements of the true instrumental profile will eliminate any approximations resulting from the Gaussian IP assumption (which arise from both wavelength calibration and convolution with the theoretical model). The Gaussian approximation used here enabled a direct comparison with an independent measurement (that also used a Gaussian IP) and is justified because any resulting systematic is unlikely to vary rapidly over the small wavelength scales for the two proximate absorption systems used \citep{Schmidt2024}.

An intriguing conclusion we can draw from the AI calculations described here are that if we require a cosmologically smooth evolution of relative isotopic abundances, such that adjacent gas clouds/galaxy envelopes are expected to exhibit the same or very similar isotopic abundances, we may have discovered real spatial fluctuations in a fundamental constant. Spatial variations of the constants of nature are a general feature of theories of gravity beyond GR e.g.\,\cite{Silva2014}, where the scale over which spatial variations occur depend on the de Broglie wavelength of the field driving $\alpha$. Spatial variations could thus be explained by theories in which $\alpha$ couples to perturbations in the matter distribution e.g.\,\cite{Barrow2001, Mota2004PLB, Mota2004MNRAS, Clifton2005}. The detailed analysis of the $z_{abs}=1.15$ HE0515$-$4414 complex by \cite{Quast2008} suggests that it may be a disrupted gaseous system associated with a merger process. That interpretation arises because (i) the $z=1.15$ complex spans $\approx\!730$ km/s. Such extreme kinematics may indicate a merger origin \citep{Prochaska2019}; (ii) there is a paucity of bright $z\approx 1.15$ galaxies detected in a shallow VLT/MUSE survey of the region (Bielby et al. 2017). Whilst this could suggest that the absorption complex does not arise in a galaxy cluster, deeper images and galaxy spectra are required to check this interpretation; (iii) neutral carbon C{\sc i} and excited levels C{\sc i}* and C{\sc i}** are detected in the right region. The relative strengths of these lines show that the right region either has a high particle density or is exposed to a very intense UV radiation field \citep{Quast2002}. These things are consistent with a merger process. Galaxy mergers are thought to enhance dark matter clumping \citep{Diemand2008}, which may be traced by the gas measured here in the $z=1.15$ complex. The Mota--Barrow model therefore potentially provides a rather natural explanation for $\alpha$ variations across this absorption complex.

Notwithstanding the above possibility, it is entirely possible that the measurements we report are a consequence of interacting galaxies, with physical conditions (particle density, ionisation fraction, metallicity, and hence isotopic abundances) that vary dramatically across the complex. An interacting scenario was developed in the detailed analysis of \cite{Quast2008}, who argue that the metallicity in the saturated (right) region is consistent with solar system abundances. If so, this may then explain our independent result that the right region returns a fine structure constant that is consistent with the terrestrial value, when using solar Mg relative isotopic abundances. In this case, our AI measurements show that dramatic isotopic abundance variations can occur over fairly small scales at early times. Models for the chemical evolution of magnesium isotopes may bifurcate at low metallicity, such that $ 0.1 \lesssim (^{25}$Mg$+^{26}$Mg$)/$Mg $\lesssim 0.6$, for ([Fe/H] $\sim$ -1.5), depending on whether enrichment from AGB and intermediate mass stars is or is not included \citep{Ashenfelter2004PRL}. However, the metallicity of the left region is not known. Further, amongst Galactic measurements, it is unclear whether such a dramatic heavy isotope spread exists empirically \citep{Vangioni2019, Kobayashi2020}. Isotope measurements at $z \approx 1$ are currently not sufficiently extensive or accurate to answer this.

If this latter interpretation is correct, the discovery of small-scale isotopic abundance variations at these redshifts suggests that Mg should no longer be used as an ``anchor'' (low $q$ value, hence insensitive to $\alpha$-variation) for varying constant studies. In the present analysis, removing Mg and re-measuring $\daa$ is not a useful exercise as we have no other anchor for the left region, and measurement errors would then permit essentially any solution. 

Parameter uncertainties have been derived, following all previous studies of this kind, by inverting the diagonal terms of the Hessian matrix. In general, caution is needed in doing so because it is possible that significant off-diagonal terms could be present, and model non-uniqueness might create multiple minima in $\chi^2$-parameter space \citep{Lee2021}. However, in this specific analysis, Figures \ref{fig:terrestrialMg} and \ref{fig:AllMg24} show no evidence for non-uniqueness. Moreover, Markov Chain Monte Carlo methods have been applied to derive parameter uncertainties, testing the simpler (and much faster) Hessian diagonal approach, finding the latter to be reliable \citep{King2009}. For these reasons, we believe parameter estimates provided in the present analysis are robust.

Whatever the interpretation, it is clear that carefully designed absorption system selection criteria will be required to derive robust $\daa$ measurements in future studies. This may include, for example, Fe{\sc ii}-only analyses, but in this case the Fe{\sc ii}1608\AA\ line will be needed to achieve good sensitivity; the Many Multiplet method was originally conceived for precisely that purpose even though its inaugural application made use of Mg{\sc ii}. An option is to use anchors other than Mg, where detected. Al{\sc ii} is ideal as there is only one stable isotope, and Si{\sc ii} is also well-suited. For DLA analyses at higher redshift, using combinations of Zn{\sc ii}, Cr{\sc ii}, Ni{\sc ii}, and Mn{\sc ii} (in conjunction with other species) may yield strong constraints, since these offer a wide range of $q$ coefficients. However, those species are only detected in damped Lyman alpha systems. In the analysis here, the right region exhibits a higher absorption component number density, causing more severe blending and larger line position errors, and hence a less accurate $\alpha$ measurement. The extent to which model non-uniqueness is generally more problematic for optically thick (saturated) systems is yet to be explored.

\section*{Acknowledgements}

Supercomputer calculations were carried out using OzSTAR at the Centre for Astrophysics and Supercomputing, Swinburne University of Technology. This research is based on observations collected at the European Southern Observatory under ESO programme 1102.A-0852 (PI: Paolo Molaro). We thank that team for making their co-added spectrum publicly available, and Professors Bob Carswell and David Mota for comments on an early draft of this paper.

\section*{Data Availability}

The ESPRESSO spectra and associated files used for this analysis are available at \url{https://doi.org/10.5281/zenodo.5512490}. The {\sc ai-vpfit} model files are available on request from the authors.

\clearpage
\bibliographystyle{mnras}
\bibliography{0515}

\bsp
\label{lastpage}
\end{document}